\documentclass[aps,prapplied,reprint,amsmath,amssymb,reprint]{revtex4-2}

\usepackage[english]{babel}
\usepackage[utf8]{inputenc}
\usepackage[colorinlistoftodos]{todonotes}
\usepackage[colorlinks,citecolor=blue,linkcolor=blue]{hyperref}
\usepackage{graphicx}	
\usepackage{bm}	
\usepackage{txfonts} 	
\usepackage[version = 4]{mhchem}     
\usepackage{siunitx}    
\usepackage{upgreek}
\usepackage{xcolor}
\usepackage{amsmath}
\usepackage[pagewise]{lineno}

\newcommand{\UIUC}{Department of Materials Science and Engineering and Materials Research Laboratory, The Grainger College of Engineering, University of Illinois at Urbana-Champaign, Illinois 61801, USA}

\begin{document}
\setlength\columnsep{25pt}

\title{Current-driven transient dynamics of magnetic-octupole domain walls in noncollinear antiferromagnets}

\author{Myoung-Woo Yoo}
\email{mwyoo@illinois.edu}
\affiliation{\UIUC}
\author{Axel Hoffmann}
\affiliation{\UIUC}

\date{\today}

\begin{abstract}
	
Current-driven magnetic-octupole domain walls in noncollinear antiferromagnets are promising for fast spintronic devices, yet their transient dynamics are not yet fully understood. Here, we investigate spin-orbit-torque-driven domain-wall motion in \ce{Mn3Sn} using a multipole micromagnetic framework. The three symmetry-related 60$^\circ$ walls exhibit equivalent current-driven dynamics, unlike their field-driven motion. For a representative 60$^\circ$ wall, the post-pulse motion is governed by relaxation of the current-induced rotation of the surrounding octupole domains, with a characteristic time of approximately 33~ps quantitatively captured by a uniform-domain analytic model. By resolving spatially nonuniform texture dynamics beyond this analytic description, the simulations further reveal an additional slow relaxation in a composite 180$^\circ$ wall associated with the relative motion of its constituent walls. The post-pulse response also varies systematically with the exchange stiffness and magnetic anisotropies. These results identify distinct domain and internal-coordinate relaxation channels and establish the intrinsic time scales governing pulse-controlled positioning of magnetic-octupole domain walls in antiferromagnetic spintronic devices.
	
\end{abstract}

\maketitle

\section{Introduction}

Current-driven domain-wall motion underpins proposed spintronic architectures for memory, logic, and unconventional computing~\cite{allwood2005, parkin2008, kumar2022, marrows2024, venkat2024, vermeulen2024}. Antiferromagnets are attractive because they offer ultrafast dynamics and weak stray fields~\cite{jungwirth2016, baltz2018, daldin2024, siddiqui2020}. However, their small net magnetization makes electrical detection and manipulation challenging. Noncollinear antiferromagnets provide a distinct opportunity because their spin textures can be represented by cluster magnetic multipoles that govern macroscopic responses such as the anomalous Hall, Nernst, magneto-optical Kerr, and spin Hall effects~\cite{suzuki2017, nakatsuji2015, higo2018, chen2021,shim2025}. In \ce{Mn3Sn}, the inverse triangular spin structure is described by a cluster magnetic-octupole order parameter, and magnetic domains correspond to different orientations of this octupole moment. Magnetic-octupole domains and domain walls have been detected through Hall and Nernst transport, directly visualized by magneto-optical imaging, and manipulated using magnetic fields and electrical currents~\cite{higo2018, reichlova2019, sugimoto2020, wu2024, yoo2024}. Recent studies have demonstrated experimentally and theoretically that magnetic-octupole domain walls in noncollinear antiferromagnets can exhibit fast current-driven motion, highlighting their potential for electrically controlled antiferromagnetic domain-wall devices~\cite{wu2024, xu2025, yamane2019, gomonay2016}.

The transient response of a domain wall after a driving force is applied or removed can reveal dynamical memory associated with the relaxation of the wall profile or other coupled degrees of freedom. In conventional ferromagnets, deformation of a moving wall can be described in terms of a D\"oring-like effective mass, giving rise to finite acceleration and deceleration times~\cite{Doring1948, Saitoh2004, Thomas2010, Torrejon2016,yoo2021}; inertial effects have also been predicted to govern switching in antiferromagnets~\cite{kimel2009}. Using a multipole micromagnetic framework, we recently showed that field-driven magnetic-octupole domain walls in \ce{Mn3Sn} deform and acquire a D\"oring-like effective mass~\cite{yoo2026}. Current-induced spin-orbit torque, however, acts simultaneously on the wall and the surrounding octupole domains. Previous theoretical work has investigated the SOT-driven dynamics of spatially uniform \ce{Mn3Sn}, including its response to current pulses~\cite{shukla2023}. How these uniform order-parameter dynamics couple to the translational motion of a spatially nonuniform texture, and how additional internal degrees of freedom modify the transient response of a composite wall, remains unresolved.

Here, we investigate the current-driven transient dynamics of magnetic-octupole domain walls in \ce{Mn3Sn} using a multipole micromagnetic model. We first show that, unlike field-driven motion, spin-orbit torque produces identical steady-state and transient dynamics for the three symmetry-related types of 60$^\circ$ walls. For a representative 60$^\circ$ wall, a dc current drives steady translational motion, whereas pulse termination produces a finite velocity relaxation and post-pulse displacement. We show that this delayed response is governed by the relaxation of the current-induced rotation of the surrounding octupole domains and that its characteristic time is consistent with the uniform-domain relaxation determined by the damping and sixfold anisotropy. We further show that a 180$^\circ$ wall behaves as a composite object consisting of three coupled 60$^\circ$ walls. Its post-pulse dynamics exhibit a fast domain-relaxation component and an additional slow mode associated with the relative separations among the constituent walls. These results reveal distinct relaxation mechanisms governing the transient motion of current-driven magnetic-octupole domain walls.

\section{Model}

\begin{figure}
	\centering
	\includegraphics[width = 8.0cm]{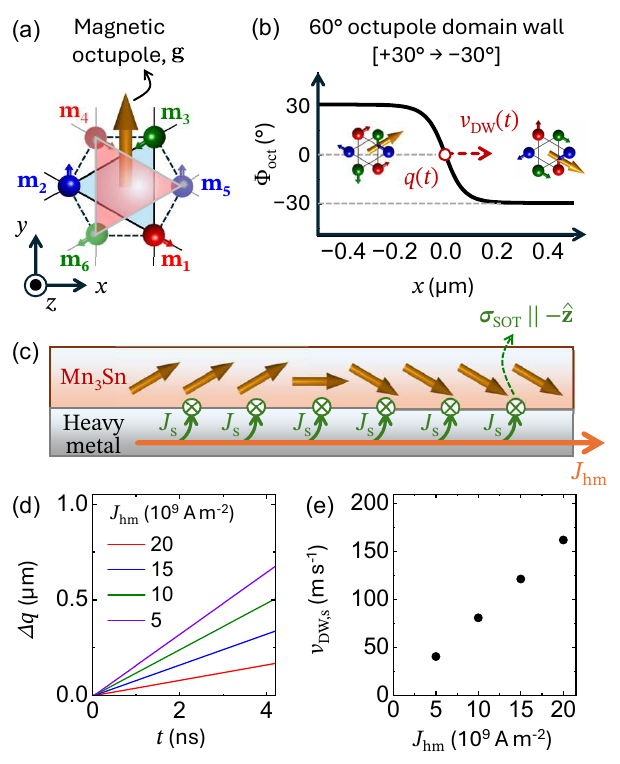}
	\caption{\label{Fig:1}
		Magnetic-octupole domain wall and dc-current-driven motion in \ce{Mn3Sn}. (a) Inverse-triangular spin structure and the corresponding cluster magnetic-octupole order parameter $\mathbf{g}$. (b) Representative equilibrium profile of a 60$^\circ$ wall connecting domains with $\Phi_{\mathrm{oct}}=+30^\circ$ and $-30^\circ$. (c) Heavy-metal/\ce{Mn3Sn} bilayer geometry, where $J_{\mathrm{hm}}$ generates a spin current $j_{\mathrm{s}}$ with $\boldsymbol{\sigma}_{\mathrm{SOT}} \parallel -\hat{\mathbf{z}}$. (d) Domain-wall displacement $\Delta q$ under different $J_{\mathrm{hm}}$. (e) Steady-state velocity $v_{\mathrm{DW,s}}$ as a function of $J_{\mathrm{hm}}$.		
	}
\end{figure}

We simulate the current-driven dynamics of magnetic-octupole domain walls using the multipole micromagnetic framework developed for \ce{Mn3Sn}~\cite{yoo2026}. In this framework, the noncollinear spin configuration within each magnetic unit cell is represented by a vector-like cluster magnetic-octupole order parameter, $\mathbf{g}$ [Fig.~\ref{Fig:1}(a)], and the corresponding coarse-grained field $\mathbf{G}(\mathbf{r},t)$ describes spatially varying octupole textures. The dynamics of $\mathbf{G}$ are calculated using the octupole equation of motion, 
\begin{equation} \label{Eq:dynamic_equation}
	\dot{\mathbf{G}} (\mathbf{r}, t) = -\gamma \mu_0 \mathbf{G} \times \mathbf{H}_{\mathrm{eff}} \quad \text{,}
\end{equation}
with an effective field,
\begin{equation} \label{Eq:effective_field}
	\mathbf{H}_{\mathrm{eff}} =	\frac{1}{6\mu_0\alpha m_{\mathrm{s}}\rho} \left(
	\frac{\partial \mathcal{E}_{\mathrm{c}}}{\partial \Phi_{\mathrm{oct}}}-\frac{\hbar \sigma_{\mathrm{SOT},z}\theta_{\mathrm{SH}}J_{\mathrm{hm}}}{2e h_{\ce{Mn3Sn}}}
	\right)	\hat{\mathbf{z}} \quad \text{,}
\end{equation}
where $\sigma_{\mathrm{SOT},z} = \boldsymbol{\sigma}_{\mathrm{SOT}} \cdot \hat{\mathbf{z}}$ is the out-of-plane component of the spin-polarization direction ($\sigma_{\mathrm{SOT},z}=-1$ in the present geometry), $\rho$ is the magnetic-unit-cell density, $\Phi_{\mathrm{oct}}$ is the in-plane orientation of the octupole order, and $\mathcal{E}_{\mathrm{c}}$ is the coarse-grained octupole energy density containing the exchange and in-plane anisotropy contributions. Here, $J_{\mathrm{hm}}$ is the in-plane charge-current density in the adjacent heavy-metal layer, and the second term represents the corresponding spin-orbit torque. Note that the Dzyaloshinskii--Moriya contribution is included in the exchange stiffness~\cite{yoo2026}.

Unless otherwise stated, we use the same material parameters as in Ref.~\cite{yoo2026}: an effective octupole magnetization of $G_0 = 6.1 \times 10^3~\mathrm{A\,m^{-1}}$, a sixfold in-plane anisotropy constant of $K_6=37.5~\mathrm{J\,m^{-3}}$, and exchange stiffness constants of $A_{\mathrm{ex},x} = A_{\mathrm{ex},y} = 7.3\times10^{-13}~\mathrm{J\,m^{-1}}$ and $A_{\mathrm{ex},z} = 5.6 \times 10^{-13}~\mathrm{J\,m^{-1}}$. Here $K_6$ denotes the full sixfold anisotropy-barrier height, such that
$\mathcal{E}_6 = (K_6/2)(1 + \cos6\Phi_{\mathrm{oct}})$. We use a Gilbert damping constant of $\alpha = 0.003$, a sublattice moment of $m_s=3\mu_B$, and a gyromagnetic ratio of $|\gamma| = 1.76\times10^{11}~\mathrm{rad\,s^{-1}\,T^{-1}}$. Note that $G_0$ is the effective magnetization associated with the octupole order and is numerically equal to the small net ferromagnetic-moment density. For the spin-orbit-torque-driven simulations, we use a spin Hall angle of $\theta_{\mathrm{SH}} = 0.1$, a \ce{Mn3Sn} thickness of $h_{\ce{Mn3Sn}} = 40~\mathrm{nm}$, and a spin polarization $\boldsymbol{\sigma}_{\mathrm{SOT}}\parallel-\hat{\mathbf{z}}$. The spin-orbit torque is treated as an effective torque averaged over the \ce{Mn3Sn} thickness, representing the net torque transferred to the exchange-coupled magnetic order without explicitly resolving its spatial variation along the film thickness.

Unless otherwise noted, the sixfold anisotropy is the only in-plane anisotropy in Sec.~\ref{Sec:60}; the twofold term is introduced only for the composite $180^\circ$ wall in Secs.~\ref{Sec:180} and \ref{Sec:internal}. The simulations are performed on a one-dimensional grid along $x$ with a cell size of $10\times10\times10~\mathrm{nm^3}$ and a total length of $20~\mathrm{\mu m}$, using periodic boundary conditions in the $y$- and $z$-directions and free boundary conditions along $x$. The wall position $q(t)$ is obtained by linear interpolation of the octupole-angle profile between neighboring cells.

\section{Current-driven motion of a 60$^\circ$ octupole domain wall}\label{Sec:60}

We first consider the three symmetry-related types of 60$^\circ$ magnetic-octupole domain walls connecting the $+90^\circ$ and $+30^\circ$, $+30^\circ$ and $-30^\circ$, and $-30^\circ$ and $-90^\circ$ domains. Our simulations show that all three walls exhibit identical steady-state and transient dynamics under the present spin-orbit-torque geometry. This equivalence follows from the symmetry of the driving term. Because the spin polarization is normal to the kagome plane, the spin-orbit torque is invariant under the transformation $\Phi_{\mathrm{oct}} \rightarrow \Phi_{\mathrm{oct}}+60^\circ$, which maps the three wall types onto one another. In contrast, an in-plane magnetic field couples to the absolute orientation of the octupole order and renders the three walls dynamically inequivalent~\cite{yoo2026}. We therefore focus below on the representative wall between the $\Phi_{\mathrm{oct}}=+30^\circ$ and $-30^\circ$ domains [Fig.~\ref{Fig:1}(b)]. In the heavy-metal/\ce{Mn3Sn} bilayer geometry, an in-plane charge-current density $J_{\mathrm{hm}}$ generates a transverse spin current $\mathbf{j}_{\mathrm{s}}$ that is injected into \ce{Mn3Sn} with spin polarization $\boldsymbol{\sigma}_{\mathrm{SOT}}\parallel-\hat{\mathbf{z}}$. The resulting spin-orbit torque drives the magnetic-octupole domain wall [Fig.~\ref{Fig:1}(c)]. The instantaneous wall position $q(t)$ is defined as the position at which the local octupole angle equals the average orientation of the two surrounding domains,
\begin{equation} \label{Eq:dw_position}
	\Phi_{\mathrm{oct}}[q(t),t]	= \overline{\Phi}_{\mathrm{oct}}(t)	= \frac{\Phi_{\mathrm{L}}(t)+\Phi_{\mathrm{R}}(t)}{2},
\end{equation}
where $\Phi_{\mathrm{L}}(t)$ and $\Phi_{\mathrm{R}}(t)$ are the octupole angles far from the wall on its left and right sides, respectively. The domain-wall velocity is defined as $v_{\mathrm{DW}}=dq/dt$.

We first examine the motion of a 60$^\circ$ magnetic-octupole domain wall under a dc current. For all current densities considered, the wall displacement $\Delta q$ increases linearly with time [Fig.~\ref{Fig:1}(d)], demonstrating steady translational motion driven by spin-orbit torque. The slope of $\Delta q(t)$ increases with $J_{\mathrm{hm}}$, and the steady-state velocity extracted from the slopes, $v_{\mathrm{DW,s}}$, increases linearly with the applied current density $J_{\mathrm{hm}}$ [Fig.~\ref{Fig:1}(e)]. All current densities considered here lie well below the threshold for continuous octupole rotation, $J_c = 6e h_{\ce{Mn3Sn}} K_6/(\hbar\theta_{\mathrm{SH}}) \simeq 1.4 \times 10^{11}~\mathrm{A\,m^{-2}}$, so that the surrounding domains remain close to their equilibrium orientations throughout and the wall moves in a steady, non-precessional regime~\cite{yoo2026}.

\begin{figure}
	\centering
	\includegraphics[width = 8.0cm]{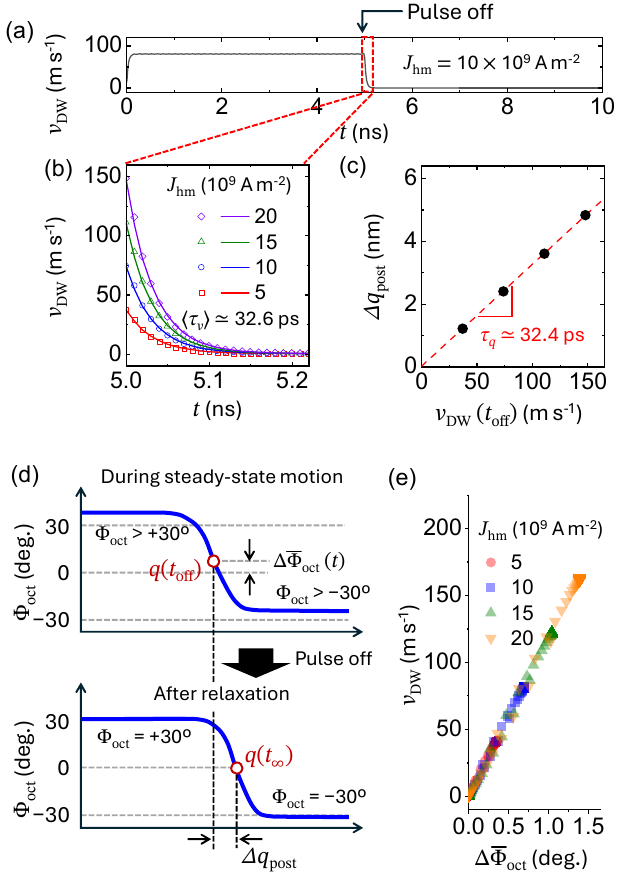}
	\caption{\label{Fig:2}
		Transient dynamics of a 60$^\circ$ magnetic-octupole domain wall under pulsed-current excitation. (a) Domain-wall velocity $v_{\mathrm{DW}}(t)$ for a 5-ns current pulse with $J_{\mathrm{hm}} = 10 \times 10^9~\mathrm{A\,m^{-2}}$. (b) Post-pulse velocity decay for several current densities, plotted as a function of $t$. Solid lines are single-exponential fits, yielding $\langle\tau_v\rangle \simeq 32.6~\mathrm{ps}$. (c) Post-pulse displacement $\Delta q_{\mathrm{post}}$ as a function of the velocity at pulse termination. The linear fit yields $\tau_q \simeq 32.4~\mathrm{ps}$. (d) Schematic octupole-angle profiles during steady-state motion and after relaxation, illustrating the residual domain rotation and the resulting post-pulse displacement. (e) Domain-wall velocity as a function of the current-induced average rotation of the surrounding domains, $\Delta\overline{\Phi}_{\mathrm{oct}}$. Data for different current densities collapse onto a single curve.		
	}
\end{figure}

Next, we investigate the transient dynamics of the 60$^\circ$ magnetic-octupole domain wall using a rectangular current pulse with a duration of 5~ns. The pulse duration is much longer than the relevant relaxation times, ensuring that the wall reaches its steady-state velocity before pulse termination. Upon pulse termination at $t_{\mathrm{off}}=5~\mathrm{ns}$, the domain-wall velocity decays over a finite time rather than vanishing immediately [Fig.~\ref{Fig:2}(a)]. The residual velocity drives continued wall motion after the current is removed, producing a finite post-pulse displacement, $\Delta q_{\mathrm{post}}$.

The velocity transients following pulse termination are well described by an exponential decay,
\begin{equation} \label{Eq:dw_velocity}
	v_{\mathrm{DW}}(t) = v_{\mathrm{DW}}(t_{\mathrm{off}})\, e^{-(t-t_{\mathrm{off}})/\tau_v} \qquad (t\geq t_{\mathrm{off}}),
\end{equation}
where $t_{\mathrm{off}}$ denotes the pulse-off time and $\tau_v$ is the velocity-relaxation time. As shown in Fig.~\ref{Fig:2}(b), this form reproduces the post-pulse velocity decay, yielding a characteristic relaxation time of $\langle\tau_v\rangle \simeq 32.6~\mathrm{ps}$ that is essentially independent of the applied current density. The same relaxation time governs the onset of the motion: the velocity rise following the leading edge of the pulse, $v_{\mathrm{DW}}(t) = v_{\mathrm{DW,s}}[1-e^{-t/\tau_v}]$, is the exact mirror of the post-pulse decay of Eq.~\ref{Eq:dw_velocity}, with the same $\tau_v$ within the fitting accuracy. For an exponential decay, the corresponding post-pulse displacement is
\begin{equation} \label{Eq:post_displacement}
	\Delta q_{\mathrm{post}} = \int_{t_{\mathrm{off}}}^{\infty} v_{\mathrm{DW}}(t)\,dt = \tau_v v_{\mathrm{DW}}(t_{\mathrm{off}}).
\end{equation}
Consistent with Eq.~\ref{Eq:post_displacement}, $\Delta q_{\mathrm{post}}$ increases linearly with the velocity at pulse termination [Fig.~\ref{Fig:2}(c)]. The slope yields $\tau_q\simeq32.4~\mathrm{ps}$, in excellent agreement with the velocity-relaxation time $\langle\tau_v\rangle \simeq32.6~\mathrm{ps}$ obtained independently from the velocity decay. This agreement provides a consistency check of the single-exponential description and confirms that integration of the residual velocity quantitatively accounts for the observed post-pulse displacement.

To identify the origin of the delayed wall motion, we examine the current-induced rotation of the octupole domains on the two sides of the wall. During current-driven motion, the orientations $\Phi_{\mathrm{L}}(t)$ and $\Phi_{\mathrm{R}}(t)$ deviate from their equilibrium values while their angular difference remains fixed at $60^\circ$ [Fig.~\ref{Fig:2}(d)]. The two domains therefore undergo a common rotation rather than changing their relative orientation. We characterize this collective domain rotation by
\begin{equation} \label{Eq:delta_Phi}
	\Delta\overline{\Phi}_{\mathrm{oct}}(t) = \overline{\Phi}_{\mathrm{oct}}(t) - \overline{\Phi}_{\mathrm{oct}}^{(0)},
\end{equation}
where $\overline{\Phi}_{\mathrm{oct}}(t)$ is the instantaneous average orientation used in Eq.~\ref{Eq:dw_position}, and $\overline{\Phi}_{\mathrm{oct}}^{(0)}$ is its equilibrium value. After pulse termination, the residual domain rotation therefore sustains the wall motion until the surrounding domains return to equilibrium, as illustrated in Fig.~\ref{Fig:2}(d).

The characteristic relaxation time of the surrounding domains can be obtained analytically from the uniform-octupole equation of motion derived in Ref.~\cite{yoo2026}. Writing $\overline{\Phi}_{\mathrm{oct}} = \overline{\Phi}_{\mathrm{oct}}^{(0)} + \delta\Phi$ and linearizing the sixfold anisotropy about an equilibrium orientation gives an overdamped relaxation, $\delta\Phi (t) \propto \exp (-t / \tau_\Phi)$, with
\begin{equation} \label{Eq:domain_relaxation}
	\tau_{\Phi} = \alpha \frac{m_s\rho}{3|\gamma|K_6},
\end{equation}
where $\rho$ is the magnetic-unit-cell density. For the parameters used here, Eq.~\ref{Eq:domain_relaxation} gives $\tau_{\Phi} \simeq 32.9~\mathrm{ps}$. The velocity-relaxation time of the domain wall, $\langle\tau_v\rangle \simeq 32.6~\mathrm{ps}$ [Fig.~\ref{Fig:2}(b)], agrees with $\tau_\Phi$. This confirms that the uniform in-plane octupole degree of freedom supports a single relaxational mode within the present reduced model. Equation~\ref{Eq:domain_relaxation} shows that the relaxation is set entirely by material parameters, $\tau_\Phi \propto \alpha/K_6$~\cite{konakanchi2025}.

This behavior follows directly from a collective-coordinate description of the wall. Writing $\Phi_{\mathrm{oct}}(x,t) = \Phi_w[x-q(t)] + \Delta\overline{\Phi}_{\mathrm{oct}}(t)$, where $\Phi_w$ is the equilibrium wall profile, the wall velocity is then set by the instantaneous domain rotation alone,
\begin{equation} \label{Eq:velocity_domain_rotation}
	v_{\mathrm{DW}} = \frac{3\delta_w}{2}\,\frac{\Delta\overline{\Phi}_{\mathrm{oct}}}{\tau_\Phi},
\end{equation}
where $\delta_w = \pi\sqrt{A_{\mathrm{ex}}}/(3\sqrt{K_6}) \simeq 146~\mathrm{nm}$ is the 60$^\circ$ wall width~\cite{yoo2026}. Note that all angles in equations are expressed in radians, whereas the figures use degrees.

Equation~\ref{Eq:velocity_domain_rotation} shows that the instantaneous wall velocity is determined by the rotation of the surrounding domains rather than directly by the applied current. The predicted slope of $116~\mathrm{m\,s^{-1}}$ per degree of domain rotation is in quantitative agreement with Fig.~\ref{Fig:2}(e). After pulse termination, the residual domain rotation decays as
$\Delta\overline{\Phi}_{\mathrm{oct}}(t)
=
\Delta\overline{\Phi}_{\mathrm{oct}}(t_{\mathrm{off}})
\exp[-(t-t_{\mathrm{off}})/\tau_\Phi]$.
Combining this relaxation with Eq.~\ref{Eq:velocity_domain_rotation} and integrating the resulting velocity gives
\begin{equation}
	\label{Eq:geometric_displacement}
	\Delta q_{\mathrm{post}}
	=
	\frac{3}{2}\delta_w
	\Delta\overline{\Phi}_{\mathrm{oct}}(t_{\mathrm{off}}).
\end{equation}
The explicit relaxation-time dependence therefore cancels. For a given residual domain rotation, the post-pulse displacement is independent of the damping, while the exchange stiffness and sixfold anisotropy enter through the wall width $\delta_w$. Equation~\ref{Eq:geometric_displacement} predicts approximately $3.8~\mathrm{nm}$ of post-pulse displacement per degree of residual domain rotation, consistent with the product of the independently obtained slopes in Figs.~\ref{Fig:2}(c) and \ref{Fig:2}(e).

\section{Current-driven motion of a 180$^\circ$ octupole domain wall}\label{Sec:180}

We next examine the dynamics of a 180$^\circ$ magnetic-octupole domain wall. Owing to the sixfold in-plane anisotropy of \ce{Mn3Sn}, a 180$^\circ$ wall forms a staircase-like structure consisting of three coupled 60$^\circ$ walls~\cite{sugimoto2020,yoo2026}. In the absence of magnetostatic energy or an additional anisotropy, however, the relative separations between the constituent walls are not uniquely determined. To stabilize the composite wall structure, we introduce an additional twofold anisotropy along the $y$ direction, as in our previous work~\cite{yoo2026},
\begin{equation}  \label{Eq:anisotropy_180}
	\mathcal{E}_{\mathrm{ani}} = \frac{K_6}{2}\left(1+\cos 6\Phi_{\mathrm{oct}}\right)	+ \frac{K_2}{2}\left(1+\cos 2\Phi_{\mathrm{oct}}\right),
\end{equation}
where $K_6$ and $K_2$ denote the sixfold and twofold anisotropy barriers, respectively. We use $K_2=10~\mathrm{J\,m^{-3}}$, corresponding to approximately 27\% of $K_6=37.5~\mathrm{J\,m^{-3}}$. Such a twofold anisotropy can be induced in \ce{Mn3Sn} by in-plane tensile strain~\cite{higo2022,ikhlas2022,yoon2023,he2024,shukla2024}.

\begin{figure}
	\centering
	\includegraphics[width = 8.0cm]{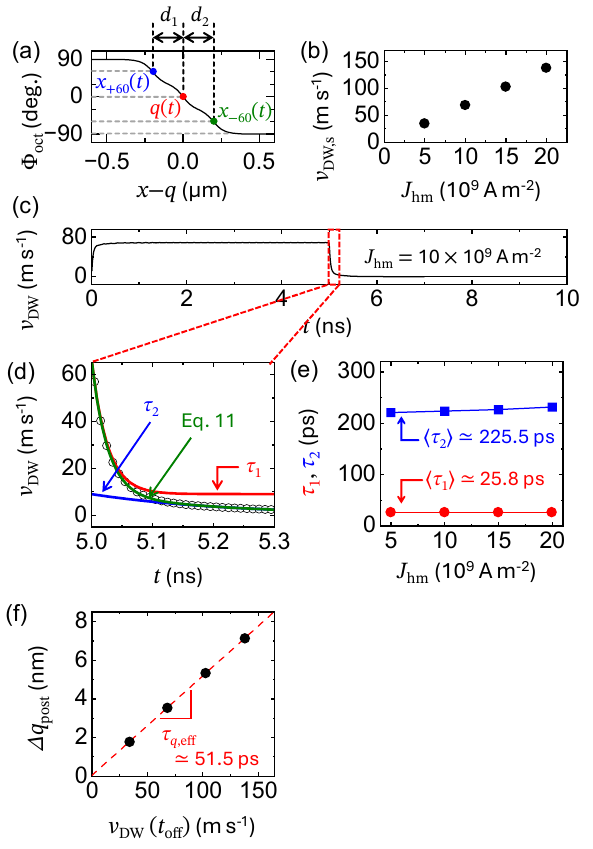}
	\caption{\label{Fig:3}
		Current-driven dynamics of a 180$^\circ$ magnetic-octupole domain wall. (a) Equilibrium octupole-angle profile of the domain wall, showing the wall position $q(t)$, the constituent-wall positions $x_{+60}(t)$ and $x_{-60}(t)$, and the internal separations $d_1$ and $d_2$. (b) Steady-state velocity $v_{\mathrm{DW,s}}$ as a function of $J_{\mathrm{hm}}$. (c) Domain-wall velocity under a 5-ns current pulse with $J_{\mathrm{hm}} = 10\times10^9~\mathrm{A\,m^{-2}}$. (d) Post-pulse velocity decay fitted by fast and slow exponential components with relaxation times $\tau_1$ and $\tau_2$. (e) Current-density dependence of $\tau_1$ and $\tau_2$, with average values of approximately $25.8~\mathrm{ps}$ and $225.5~\mathrm{ps}$, respectively. (f) Post-pulse displacement $\Delta q_{\mathrm{post}}$ as a function of $v_{\mathrm{DW}} (t_{\mathrm{off}})$. The linear fit yields $\tau_{q,\mathrm{eff}} \simeq 51.5~\mathrm{ps}$.		
	}
\end{figure}

The equilibrium profile of the resulting 180$^\circ$ wall is shown in Fig.~\ref{Fig:3}(a). The wall position $q(t)$ is defined using Eq.~\ref{Eq:dw_position}, with $\Phi_L$ and $\Phi_R$ taken from the far-left and far-right domains, respectively. Under a dc current, the composite wall undergoes steady translational motion, and its steady-state velocity $v_{\mathrm{DW,s}}$ increases with $J_{\mathrm{hm}}$ [Fig.~\ref{Fig:3}(b)].

We next apply a rectangular current pulse with a duration of 5~ns. During the pulse, the wall velocity approaches a steady-state value, whereas after pulse termination it decays gradually toward zero [Fig.~\ref{Fig:3}(c)]. In contrast to the single-exponential relaxation of the 60$^\circ$ wall, the post-pulse velocity of the 180$^\circ$ wall cannot be described by a single relaxation time. Instead, it is well reproduced by a sum of two exponential components,
\begin{equation} \label{Eq:180_velocity}
	v_{\mathrm{DW}}(t) = A_1 e^{-(t-t_{\mathrm{off}})/\tau_1} + A_2 e^{-(t-t_{\mathrm{off}})/\tau_2},
\end{equation}
for $t \geq t_{\mathrm{off}}$, where $A_1$ and $A_2$ are the amplitudes of the fast and slow components, respectively, and $\tau_1$ and $\tau_2$ are their corresponding relaxation times. As shown in Fig.~\ref{Fig:3}(d), the fast component governs the initial velocity decay, while the slow component produces a longer relaxation tail.

Both relaxation times depend only weakly on the applied current density [Fig.~\ref{Fig:3}(e)]. Their average values are $\langle \tau_1 \rangle \simeq 25.8~\mathrm{ps}$ and $\langle \tau_2 \rangle \simeq 225.5~\mathrm{ps}$. The fast relaxation time is of the same order as the velocity-relaxation time of the 60$^\circ$ wall, but it is about 20\% shorter than the uniform-domain value $\tau_\Phi \simeq 32~\mathrm{ps}$. The slow time scale is nearly one order of magnitude larger. Since the uniform in-plane octupole coordinate supports only a single relaxation mode within the present model, the appearance of two time scales, together with the shift of the faster one away from $\tau_\Phi$, indicates that the translational motion of the composite wall is coupled to additional internal degrees of freedom that are absent for an isolated 60$^\circ$ wall.

For the double-exponential relaxation, the post-pulse displacement is given by
\begin{equation} \label{Eq:180_post_displacement}
	\Delta q_{\mathrm{post}} = \int_{t_{\mathrm{off}}}^{\infty} v_{\mathrm{DW}}(t)\,dt = A_1\tau_1+A_2\tau_2.
\end{equation}
Because the velocity at pulse termination is
$v_{\mathrm{DW}} (t_{\mathrm{off}}) = A_1 + A_2$,
an effective displacement-relaxation time can be defined as
\begin{equation} \label{Eq:tau_q_eff}
	\tau_{q,\mathrm{eff}} = \frac{A_1\tau_1+A_2\tau_2}{A_1+A_2},
\end{equation}
such that $\Delta q_{\mathrm{post}} = \tau_{q,\mathrm{eff}} v_{\mathrm{DW}}(t_{\mathrm{off}})$,
with $\langle \tau_{q,\mathrm{eff}} \rangle \simeq 52.4~\mathrm{ps}$ from the fitted amplitudes.
Consistent with this relation, $\Delta q_{\mathrm{post}}$ increases linearly with the velocity at pulse termination [Fig.~\ref{Fig:3}(f)], and the slope yields $\tau_{q,\mathrm{eff}} \simeq 51.5~\mathrm{ps}$.
The two values of $\tau_{q,\mathrm{eff}}$ are considerably larger than $\tau_1$: although the slow component carries only about 14\% of the velocity at pulse termination, its nearly ninefold longer relaxation time makes it responsible for more than half of the total post-pulse displacement.

\begin{figure}
	\centering
	\includegraphics[width = 8.0cm]{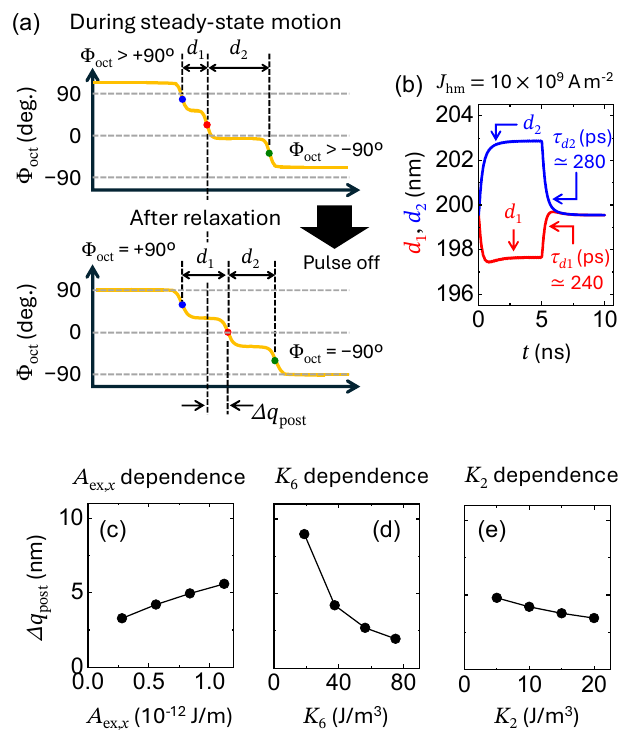}
	\caption{\label{Fig:4}
		Internal separation dynamics and tunability of the post-pulse motion of a 180$^\circ$ magnetic-octupole domain wall. (a) Octupole-angle profiles during steady-state motion and after relaxation following pulse termination. The internal separations are defined as $d_1 = q - x_{+60}$ and $d_2 = x_{-60} - q$. During current-driven motion, $d_1$ and $d_2$ become unequal; after pulse termination, they return toward equilibrium while the wall undergoes an additional displacement $\Delta q_{\mathrm{post}}$. (b) Time evolution of $d_1$ and $d_2$ for $J_{\mathrm{hm}} = 10 \times 10^9~\mathrm{A\,m^{-2}}$, with relaxation times $\tau_{d_1} \simeq 240~\mathrm{ps}$ and $\tau_{d_2} \simeq 280~\mathrm{ps}$. (c)--(e) Dependence of $\Delta q_{\mathrm{post}}$ on $A_{\mathrm{ex},x}$, $K_6$, and $K_2$, respectively.
	}
\end{figure}

\section{Internal dynamics of a 180$^\circ$ domain wall}\label{Sec:internal}

To investigate the internal dynamics accompanying the slow relaxation of the 180$^\circ$ wall, we track the relative positions of its three constituent 60$^\circ$ walls. As shown in Fig.~\ref{Fig:3}(a), we define $x_{+60}(t)$ and $x_{-60}(t)$ as the positions satisfying
\begin{equation} \label{Eq:x+60_x-60}
	\Phi_{\mathrm{oct}}\!\left[x_{\pm 60}(t),t\right] = \Phi_{\mathrm{oct}}\!\left[q(t),t\right]\pm 60^\circ ,
\end{equation}
respectively, where $q(t)$ is the overall wall position defined by Eq.~\ref{Eq:dw_position}. The two internal separations are then defined as
\begin{equation} \label{Eq:internal_separations}
	\begin{aligned}
		d_1(t) &= q(t)-x_{+60}(t), \\
		d_2(t) &= x_{-60}(t)-q(t).
	\end{aligned}
\end{equation}
These quantities characterize the relative separations among the three constituent 60$^\circ$ walls while remaining insensitive to a common rotation of the surrounding octupole domains.

Figure~\ref{Fig:4}(a) schematically compares the octupole-angle profiles during steady-state motion and after relaxation following pulse termination. During current-driven motion, $d_1$ and $d_2$ become unequal, indicating an asymmetric deformation of the domain wall. Because isolated 60$^\circ$ domain walls exhibit equivalent current-driven motion, this asymmetry suggests that the additional twofold anisotropy and the mutual coupling among the three constituent walls govern the internal deformation of the composite wall. After the current is removed, the two separations return toward their common equilibrium value while the wall undergoes an additional displacement $\Delta q_{\mathrm{post}}$.

The time evolutions of $d_1$ and $d_2$ are shown in Fig.~\ref{Fig:4}(b) for $J_{\mathrm{hm}}=10\times10^9~\mathrm{A\,m^{-2}}$. During the current pulse, $d_1$ decreases while $d_2$ increases from their equilibrium values. Following pulse termination at $t_{\mathrm{off}}=5~\mathrm{ns}$, both separations relax toward equilibrium, with characteristic time scales of $\tau_{d_1}\simeq 240 ~\mathrm{ps}$ and $\tau_{d_2} \simeq 280~\mathrm{ps}$. These time scales are approximately one order of magnitude longer than the fast velocity-relaxation time $\tau_1$ and are close to the slow component $\tau_2 \simeq 230~\mathrm{ps}$. Together with the single relaxation mode of the uniform in-plane octupole coordinate, this correspondence supports the assignment of the fast and slow velocity components predominantly to the domain-rotation and antisymmetric internal-separation channels, respectively. The coupling of these two relaxation channels accounts for the double-exponential velocity response in Eq.~\ref{Eq:180_velocity}.

We finally examine the parameter dependence of the post-pulse motion of the 180$^\circ$ wall by varying $A_{\mathrm{ex},x}$, $K_6$, and $K_2$ while keeping the pulse conditions and all other independent parameters fixed. Figures~\ref{Fig:4}(c)--\ref{Fig:4}(e) show that $\Delta q_{\mathrm{post}}$ increases with $A_{\mathrm{ex},x}$ but decreases with $K_6$ and $K_2$, with a particularly strong dependence on the sixfold anisotropy. The Dzyaloshinskii--Moriya contribution is incorporated into $A_{\mathrm{ex},x}$ following Ref.~\cite{yoo2026}. These parameter dependences reflect both the relaxation of the surrounding domains and changes in the internal structure of the composite wall. Although the $K_2$ dependence is weaker than that of $K_6$, its effect on $\Delta q_{\mathrm{post}}$ cannot be accounted for solely by the small change in the uniform-domain relaxation, suggesting that changes in the widths and overlap of the constituent walls also contribute. This interpretation is consistent with the previously reported reduction in the total width of the 180$^\circ$ wall with increasing $K_2$~\cite{yoo2026}.

The relaxation channels identified above set the time scale over which a wall
settles into a well-defined position after a current pulse. Defining the settling time as the time required to complete 95\% of $\Delta q_{\mathrm{post}}$, the 60$^\circ$ wall reaches within 5\% of its final post-pulse position in approximately $100~\mathrm{ps}$. For the representative 180$^\circ$ wall at $J_{\mathrm{hm}}=10\times10^9~\mathrm{A\,m^{-2}}$, the fitted amplitudes and relaxation times in Eq.~\ref{Eq:180_velocity} give a settling time of approximately 550~ps. The internal degrees of freedom of the composite wall therefore increase the post-pulse settling time by a factor of about five, providing an intrinsic time scale relevant to pulse-controlled positioning and subsequent position readout. The strain-induced twofold anisotropy provides an experimentally accessible means of tuning the post-pulse displacement.

\section{Conclusion}

We have investigated the current-driven transient dynamics of magnetic-octupole domain walls in \ce{Mn3Sn} using a multipole micromagnetic framework. For a 60$^\circ$ wall, the post-pulse motion is governed by the relaxation of the current-induced rotation of the surrounding domains, with a characteristic time quantitatively reproduced by a uniform-domain analytic model. For a composite 180$^\circ$ wall, the simulations reveal an additional slow relaxation associated with the relative separations among its constituent walls, which substantially extends the time required for the wall to reach its final post-pulse position. The post-pulse response also varies systematically with the exchange stiffness and magnetic anisotropies, with the strain-induced twofold anisotropy providing an experimentally accessible means of tuning the post-pulse displacement. These results identify distinct domain and internal-coordinate relaxation channels and establish the intrinsic time scales governing pulse-controlled positioning and subsequent readout of magnetic-octupole domain walls in antiferromagnetic spintronic devices.

\section{Acknowledgements}

This work was supported by the National Science Foundation (NSF) through the Center for Advanced Semiconductor Chips with Accelerated Performance (ASAP) Industry-University Cooperative Research Center under NSF Cooperative Agreement No. EEC-2231625.

\section{Data availability}

The data that support the findings of this study are available from the corresponding author upon reasonable request.

\bibliography{references}{}

@article{allwood2005,
  title = {Magnetic {{Domain-Wall Logic}}},
  author = {Allwood, D. A. and Xiong, G. and Faulkner, C. C. and Atkinson, D. and Petit, D. and Cowburn, R. P.},
  year = 2005,
  month = sep,
  journal = {Science},
  volume = {309},
  number = {5741},
  pages = {1688--1692},
  issn = {0036-8075, 1095-9203},
  doi = {10.1126/science.1108813},
  urldate = {2026-07-13}
}

@article{kumar2022,
  title = {Domain Wall Memory: {{Physics}}, Materials, and Devices},
  author = {Kumar, D. and Jin, T. and Sbiaa, R. and Kl{\"a}ui, M. and Bedanta, S.
            and Fukami, S. and Ravelosona, D. and Yang, S.-H. and Liu, X.
            and Piramanayagam, S. N.},
  year = 2022,
  month = may,
  journal = {Physics Reports},
  volume = {958},
  pages = {1--35},
  issn = {03701573},
  doi = {10.1016/j.physrep.2022.02.001},
  urldate = {2026-07-13}
}

@article{marrows2024,
  title = {Neuromorphic Computing with Spintronics},
  author = {Marrows, C. H. and Barker, J. and Moore, T. A. and Moorsom, T.},
  year = 2024,
  month = apr,
  journal = {npj Spintronics},
  volume = {2},
  number = {1},
  pages = {12},
  issn = {2948-2119},
  doi = {10.1038/s44306-024-00019-2},
  urldate = {2026-07-13}
}

@article{parkin2008,
  title = {Magnetic {{Domain-Wall Racetrack Memory}}},
  author = {Parkin, Stuart S. P. and Hayashi, Masamitsu and Thomas, Luc},
  year = 2008,
  month = apr,
  journal = {Science},
  volume = {320},
  number = {5873},
  pages = {190--194},
  issn = {0036-8075, 1095-9203},
  doi = {10.1126/science.1145799},
  urldate = {2026-07-13}
}

@article{venkat2024,
  title = {Magnetic Domain Walls: Types, Processes and Applications},
  author = {Venkat, G and Allwood, D A and Hayward, T J},
  year = 2024,
  month = feb,
  journal = {Journal of Physics D: Applied Physics},
  volume = {57},
  number = {6},
  pages = {063001},
  issn = {0022-3727, 1361-6463},
  doi = {10.1088/1361-6463/ad0568},
  urldate = {2026-07-13}
}

@article{vermeulen2024,
  title = {Progress in {{Spin Logic Devices Based}} on {{Domain-Wall Motion}}},
  author = {Vermeulen, Bob Bert and Sor{\'e}e, Bart and Couet, Sebastien and Temst, Kristiaan and Nguyen, Van Dai},
  year = 2024,
  month = may,
  journal = {Micromachines},
  volume = {15},
  number = {6},
  pages = {696},
  issn = {2072-666X},
  doi = {10.3390/mi15060696},
  urldate = {2026-07-13}
}

@article{baltz2018,
  title = {Antiferromagnetic Spintronics},
  author = {Baltz, V. and Manchon, A. and Tsoi, M. and Moriyama, T. and Ono, T. and Tserkovnyak, Y.},
  year = 2018,
  month = feb,
  journal = {Reviews of Modern Physics},
  volume = {90},
  number = {1},
  pages = {015005},
  issn = {0034-6861, 1539-0756},
  doi = {10.1103/RevModPhys.90.015005},
  urldate = {2026-07-13}
}

@article{daldin2024,
  title = {Antiferromagnetic Spintronics and Beyond},
  author = {Dal Din, A. and Amin, O. J. and Wadley, P. and Edmonds, K. W.},
  year = 2024,
  month = jul,
  journal = {npj Spintronics},
  volume = {2},
  number = {1},
  pages = {25},
  issn = {2948-2119},
  doi = {10.1038/s44306-024-00029-0},
  urldate = {2026-07-13}
}

@article{jungwirth2016,
  title = {Antiferromagnetic Spintronics},
  author = {Jungwirth, T. and Marti, X. and Wadley, P. and Wunderlich, J.},
  year = 2016,
  month = mar,
  journal = {Nature Nanotechnology},
  volume = {11},
  number = {3},
  pages = {231--241},
  issn = {1748-3387, 1748-3395},
  doi = {10.1038/nnano.2016.18},
  urldate = {2026-07-13}
}

@article{siddiqui2020,
  title = {Metallic Antiferromagnets},
  author = {Siddiqui, Saima A. and Sklenar, Joseph and Kang, Kisung and Gilbert, Matthew J. and Schleife, Andr{\'e} and Mason, Nadya and Hoffmann, Axel},
  year = 2020,
  month = jul,
  journal = {Journal of Applied Physics},
  volume = {128},
  number = {4},
  pages = {040904},
  issn = {0021-8979, 1089-7550},
  doi = {10.1063/5.0009445},
  urldate = {2026-07-13}
}

@article{chen2021,
  title = {Anomalous Transport Due to {{Weyl}} Fermions in the Chiral Antiferromagnets {{Mn$_3$X}}, {{X}} = {{Sn}}, {{Ge}}},
  author = {Chen, Taishi and Tomita, Takahiro and Minami, Susumu and Fu, Mingxuan and Koretsune, Takashi and Kitatani, Motoharu and Muhammad, Ikhlas and {Nishio-Hamane}, Daisuke and Ishii, Rieko and Ishii, Fumiyuki and Arita, Ryotaro and Nakatsuji, Satoru},
  year = 2021,
  month = dec,
  journal = {Nature Communications},
  volume = {12},
  number = {1},
  pages = {572},
  issn = {2041-1723},
  doi = {10.1038/s41467-020-20838-1},
  urldate = {2021-11-29}
}

@article{higo2018,
  title = {Large Magneto-Optical {{Kerr}} Effect and Imaging of Magnetic Octupole Domains in an Antiferromagnetic Metal},
  author = {Higo, Tomoya and Man, Huiyuan and Gopman, Daniel B. and Wu, Liang and Koretsune, Takashi and {van 't Erve}, Olaf M. J. and Kabanov, Yury P. and Rees, Dylan and Li, Yufan and Suzuki, Michi-To and Patankar, Shreyas and Ikhlas, Muhammad and Chien, C. L. and Arita, Ryotaro and Shull, Robert D. and Orenstein, Joseph and Nakatsuji, Satoru},
  year = 2018,
  month = feb,
  journal = {Nature Photonics},
  volume = {12},
  number = {2},
  pages = {73--78},
  issn = {1749-4885, 1749-4893},
  doi = {10.1038/s41566-017-0086-z},
  urldate = {2021-11-30}
}

@article{nakatsuji2015,
  title = {Large Anomalous {{Hall}} Effect in a Non-Collinear Antiferromagnet at Room Temperature},
  author = {Nakatsuji, Satoru and Kiyohara, Naoki and Higo, Tomoya},
  year = 2015,
  month = nov,
  journal = {Nature},
  volume = {527},
  number = {7577},
  pages = {212--215},
  issn = {0028-0836, 1476-4687},
  doi = {10.1038/nature15723},
  urldate = {2026-07-13}
}

@article{suzuki2017,
  title = {Cluster Multipole Theory for Anomalous {{Hall}} Effect in Antiferromagnets},
  author = {Suzuki, M.-T. and Koretsune, T. and Ochi, M. and Arita, R.},
  year = 2017,
  month = mar,
  journal = {Physical Review B},
  volume = {95},
  number = {9},
  pages = {094406},
  issn = {2469-9950, 2469-9969},
  doi = {10.1103/PhysRevB.95.094406},
  urldate = {2026-07-13},
  copyright = {https://link.aps.org/licenses/aps-default-license}
}

@article{reichlova2019,
  title = {Imaging and Writing Magnetic Domains in the Non-Collinear Antiferromagnet {{Mn$_3$Sn}}},
  author = {Reichlova, Helena and Janda, Tomas and Godinho, Joao and Markou, Anastasios and Kriegner, Dominik and Schlitz, Richard and Zelezny, Jakub and Soban, Zbynek and Bejarano, Mauricio and Schultheiss, Helmut and Nemec, Petr and Jungwirth, Tomas and Felser, Claudia and Wunderlich, Joerg and Goennenwein, Sebastian T. B.},
  year = 2019,
  month = nov,
  journal = {Nature Communications},
  volume = {10},
  number = {1},
  pages = {5459},
  issn = {2041-1723},
  doi = {10.1038/s41467-019-13391-z},
  urldate = {2026-07-13}
}

@article{sugimoto2020,
  title = {Electrical Nucleation, Displacement, and Detection of Antiferromagnetic Domain Walls in the Chiral Antiferromagnet {{Mn$_3$Sn}}},
  author = {Sugimoto, Satoshi and Nakatani, Yoshinobu and Yamane, Yuta and Ikhlas, Muhammad and Kondou, Kouta and Kimata, Motoi and Tomita, Takahiro and Nakatsuji, Satoru and Otani, Yoshichika},
  year = 2020,
  month = jun,
  journal = {Communications Physics},
  volume = {3},
  number = {1},
  pages = {111},
  issn = {2399-3650},
  doi = {10.1038/s42005-020-0381-8},
  urldate = {2026-07-13}
}

@article{yoo2024,
  title = {Thermal Contribution to Current-Driven Antiferromagnetic-Order Switching},
  author = {Yoo, Myoung-Woo and Lorenz, Virginia O. and Hoffmann, Axel and Cahill, David G.},
  year = 2024,
  month = aug,
  journal = {APL Materials},
  volume = {12},
  number = {8},
  pages = {081107},
  issn = {2166-532X},
  doi = {10.1063/5.0219729},
  urldate = {2026-07-13}
}

@article{wu2024,
  title = {Current-Driven Fast Magnetic Octupole Domain-Wall Motion in Noncollinear Antiferromagnets},
  author = {Wu, Mingxing and Chen, Taishi and Nomoto, Takuya and Tserkovnyak, Yaroslav and Isshiki, Hironari and Nakatani, Yoshinobu and Higo, Tomoya and Tomita, Takahiro and Kondou, Kouta and Arita, Ryotaro and Nakatsuji, Satoru and Otani, Yoshichika},
  year = 2024,
  month = jun,
  journal = {Nature Communications},
  volume = {15},
  number = {1},
  pages = {4305},
  issn = {2041-1723},
  doi = {10.1038/s41467-024-48440-9},
  urldate = {2026-07-13}
}

@article{xu2025,
  title = {Spin-Orbit Torque-Driven Chiral Domain Wall Motion in {{Mn$_3$Sn}}},
  author = {Xu, Zhengde and Zhou, Yue and Zhang, Xue and Qiao, Yixiao and Xu, Zhuo and Shao, Dingfu and Zhu, Zhifeng},
  year = 2025,
  month = aug,
  journal = {Applied Physics Letters},
  volume = {127},
  number = {7},
  pages = {072402},
  issn = {0003-6951, 1077-3118},
  doi = {10.1063/5.0275467},
  urldate = {2026-07-13}
}

@article{doring1948,
  title = {{\"U}ber die {Tr{\"a}gheit} der {W{\"a}nde} zwischen {Wei{\ss}schen} {Bezirken}},
  author  = {D{\"o}ring, W.},
  year = 1948,
  month = jul,
  journal = {Zeitschrift f{\"u}r Naturforschung A},
  volume = {3},
  number = {7},
  pages = {373--379},
  issn = {1865-7109, 0932-0784},
  doi = {10.1515/zna-1948-0701},
  urldate = {2026-07-13},
  copyright = {http://creativecommons.org/licenses/by-nc-nd/3.0/}
}

@article{saitoh2004,
  title = {Current-Induced Resonance and Mass Determination of a Single Magnetic Domain Wall},
  author = {Saitoh, Eiji and Miyajima, Hideki and Yamaoka, Takehiro and Tatara, Gen},
  year = 2004,
  month = nov,
  journal = {Nature},
  volume = {432},
  number = {7014},
  pages = {203--206},
  issn = {0028-0836, 1476-4687},
  doi = {10.1038/nature03009},
  urldate = {2026-07-13},
  copyright = {http://www.springer.com/tdm}
}

@article{thomas2010,
  title = {Dynamics of {{Magnetic Domain Walls Under Their Own Inertia}}},
  author = {Thomas, Luc and Moriya, Rai and Rettner, Charles and Parkin, Stuart S.P.},
  year = 2010,
  month = dec,
  journal = {Science},
  volume = {330},
  number = {6012},
  pages = {1810--1813},
  issn = {0036-8075, 1095-9203},
  doi = {10.1126/science.1197468},
  urldate = {2026-07-13}
}

@article{torrejon2016,
  title = {Tunable Inertia of Chiral Magnetic Domain Walls},
  author = {Torrejon, Jacob and Martinez, Eduardo and Hayashi, Masamitsu},
  year = 2016,
  month = nov,
  journal = {Nature Communications},
  volume = {7},
  number = {1},
  pages = {13533},
  issn = {2041-1723},
  doi = {10.1038/ncomms13533},
  urldate = {2026-07-13}
}

@article{yoo2026,
  title = {Micromagnetic Simulations for Magnetic Multipoles},
  author = {Yoo, Myoung-Woo and Winkler, Roland and Hoffmann, Axel},
  year = 2026,
  month = jun,
  journal = {Applied Physics Reviews},
  volume = {13},
  number = {2},
  pages = {021423},
  issn = {1931-9401},
  doi = {10.1063/5.0302867},
  urldate = {2026-06-25}
}

@article{higo2022,
  title = {Perpendicular Full Switching of Chiral Antiferromagnetic Order by Current},
  author = {Higo, Tomoya and Kondou, Kouta and Nomoto, Takuya and Shiga, Masanobu and Sakamoto, Shoya and Chen, Xianzhe and {Nishio-Hamane}, Daisuke and Arita, Ryotaro and Otani, Yoshichika and Miwa, Shinji and Nakatsuji, Satoru},
  year = 2022,
  month = jul,
  journal = {Nature},
  volume = {607},
  number = {7919},
  pages = {474--479},
  issn = {0028-0836, 1476-4687},
  doi = {10.1038/s41586-022-04864-1},
  urldate = {2022-07-25}
}

@article{ikhlas2022,
  title = {Piezomagnetic Switching of the Anomalous {{Hall}} Effect in an Antiferromagnet at Room Temperature},
  author = {Ikhlas, M. and Dasgupta, S. and Theuss, F. and Higo, T. and Kittaka, Shunichiro and Ramshaw, B. J. and Tchernyshyov, O. and Hicks, C. W. and Nakatsuji, S.},
  year = 2022,
  month = aug,
  journal = {Nature Physics},
  volume  = {18},
  pages   = {1086--1091},
  issn = {1745-2473, 1745-2481},
  doi = {10.1038/s41567-022-01645-5},
  urldate = {2022-08-22}
}

@article{yoon2023,
  title = {Handedness Anomaly in a Non-Collinear Antiferromagnet under Spin--Orbit Torque},
  author = {Yoon, Ju-Young and Zhang, Pengxiang and Chou, Chung-Tao and Takeuchi, Yutaro and Uchimura, Tomohiro and Hou, Justin T. and Han, Jiahao and Kanai, Shun and Ohno, Hideo and Fukami, Shunsuke and Liu, Luqiao},
  year = 2023,
  month = aug,
  journal = {Nature Materials},
  volume  = {22},
  pages   = {1106--1113},
  issn = {1476-1122, 1476-4660},
  doi = {10.1038/s41563-023-01620-2},
  urldate = {2023-08-08}
}

@article{yamane2019,
  title = {Dynamics of noncollinear antiferromagnetic textures driven by spin current injection},
  author = {Yamane, Y. and Gomonay, O. and Sinova, J.},
  journal = {Physical Review B}, volume = {100}, pages = {054415}, year = {2019}
}

@article{konakanchi2025,
  title = {Electrically tunable picosecond-scale octupole fluctuations in chiral antiferromagnets},
  author = {Konakanchi, S. T. and Banerjee, S. and Rahman, M. M. and Yamane, Y.
            and Kanai, S. and Fukami, S. and Upadhyaya, P.},
  journal = {Physical Review Letters}, volume = {135}, pages = {136704}, year = {2025}
}

@article{gomonay2016,
  title = {High antiferromagnetic domain wall velocity induced by
           {N}\'{e}el spin-orbit torques},
  author = {Gomonay, O. and Jungwirth, T. and Sinova, J.},
  journal = {Physical Review Letters}, volume = {117}, pages = {017202}, year = {2016}
}

@article{kimel2009,
  title = {Inertia-driven spin switching in antiferromagnets},
  author = {Kimel, A. V. and Ivanov, B. A. and Pisarev, R. V. and Usachev, P. A.
            and Kirilyuk, A. and Rasing, T.},
  journal = {Nature Physics}, volume = {5}, pages = {727--731}, year = {2009}
}

@article{he2024,
  title = {Magnetic dynamics of strained non-collinear antiferromagnet},
  author = {He, Z. and Liu, L.},
  journal = {Journal of Applied Physics}, volume = {135}, pages = {093902}, year = {2024}
}

@article{shim2025,
  title = {Spin-{{Polarized Antiferromagnetic Metals}}},
  author = {Shim, Soho and Mehraeen, M. and Sklenar, Joseph and Zhang, Steven S.-L. and Hoffmann, Axel and Mason, Nadya},
  year = 2025,
  month = mar,
  journal = {Annual Review of Condensed Matter Physics},
  volume = {16},
  number = {1},
  pages = {103--120},
  issn = {1947-5454, 1947-5462},
  doi = {10.1146/annurev-conmatphys-042924-123620},
  urldate = {2026-08-13},
  copyright = {http://creativecommons.org/licenses/by/4.0/}
}

@article{shukla2023,
  title = {Order Parameter Dynamics in {{Mn$_3$Sn}} Driven by {{DC}} and Pulsed Spin--Orbit Torques},
  author = {Shukla, Ankit and Qian, Siyuan and Rakheja, Shaloo},
  year = 2023,
  month = sep,
  journal = {APL Materials},
  volume = {11},
  number = {9},
  pages = {091110},
  issn = {2166-532X},
  doi = {10.1063/5.0158164},
  urldate = {2026-08-17}
}

@article{shukla2024,
  title = {Impact of Strain on the {{SOT-driven}} Dynamics of Thin Film {{Mn$_3$Sn}}},
  author = {Shukla, Ankit and Qian, Siyuan and Rakheja, Shaloo},
  year = 2024,
  month = mar,
  journal = {Journal of Applied Physics},
  volume = {135},
  number = {12},
  pages = {123903},
  issn = {0021-8979, 1089-7550},
  doi = {10.1063/5.0179669},
  urldate = {2026-08-17}
}

@article{yoo2021,
  title = {Analytical Model of the Deformation-Induced Inertial Dynamics of a Magnetic Vortex},
  author = {Yoo, Myoung-Woo and Mineo, Francesca and Kim, Joo-Von},
  year = 2021,
  month = feb,
  journal = {Journal of Applied Physics},
  volume = {129},
  number = {5},
  pages = {053903},
  issn = {0021-8979, 1089-7550},
  doi = {10.1063/5.0039535},
  urldate = {2021-11-08}
}
\end{document}